\documentstyle[aps,epsf]{revtex}  

\begin{document}        

\baselineskip 14pt
\title{ Measuring the Higgs Boson Yukawa Couplings at an NLC}
\author{S. Dawson }  
\address{Physics Department, Brookhaven National Laboratory,
Upton, N.Y.~~  11973 }
\author{L.~Reina } 
\address{Physics Department, Florida State University,
 Tallahassee, FL.~~32306}
\maketitle              

\begin{abstract}        
We investigate the inclusive production of the Higgs
boson with a heavy quark pair, $t {\overline t}$ or $b
{\overline b}$, in $e^+e^-$ collisions at
high energy.  The ${\cal O}(\alpha_s)$ QCD corrections are
included.  
\
\end{abstract}   	

\def\beq{\begin{equation}}
\def\eeq{\end{equation}}
\def\beqn{\begin{eqnarray}}
\def\eeqn{\end{eqnarray}}
\def\bea{\begin{eqnarray}}
\def\eea{\end{eqnarray}} 
\def\be{\begin{equation}}
\def\ee{\end{equation}}

\section{Introduction}
\label{sec:intro}

The search for the Higgs boson is one of the primary  objectives
of present and future colliders. 
Once the Higgs boson has been discovered, it will be important to
measure its couplings to fermions and to gauge bosons.  These couplings
are completely determined in the Standard Model and 
the process $e^+e^-\rightarrow t {\overline t} h$ provides a direct
mechanism for measuring the $t {\overline t} h$ Yukawa coupling.
Since this coupling can be significantly different in a supersymmetric
model from that in the Standard Model, the measurement would provide a
means of discriminating between different models.

The associated production of a Higgs boson with a top quark pair
in $e^+e^-$ collisions  has a small rate, around $1~fb$
for $\sqrt{s}=500~GeV$ and $M_h\sim 100~GeV$.  However, the signature, 
$e^+e^-\rightarrow t {\overline t}h\rightarrow W^+W^- b {\overline b}
b {\overline b}$ is distinctive
and a precise measurement may be possible.   

A similar reaction in the $b$
quark system, $e^+e^-\rightarrow b {\overline b}h$,  is
suppressed in the Standard Model due to the
smallness of the  $b {\overline b}h$ Yukawa coupling.  In a supersymmetric 
model, however, this coupling can be enhanced for large
values of the parameter  $\tan\beta$.  In
addition,  a supersymmetric
 model contains resonant contributions not present
in the Standard Model  such as, for example,  the
process $e^+e^-\rightarrow A^0 h^0_i, A^0\rightarrow b {\overline b}$.

In order to extract the Yukawa couplings, precise predictions
for the rates, including QCD corrections, are necessary.  
The QCD corrections to the associated production of a Higgs boson with
a heavy quark pair have been computed by
two groups\cite{dr1,ditt1} and  are the subject of this note.

\section{Associated Higgs-top quark production in the Standard Model}

The Standard Model
cross section for $e^+e^-\rightarrow t\bar t h$ occurs through
both $s-$ channel photon and $s-$ channel
 $Z$ exchange.\cite{gounaris,djouadi}. 
The most relevant
contributions are those in which the Higgs boson is emitted from a top
quark leg, which are directly proportional to the $t {\overline t}h$
Yukawa coupling.
 The contribution when  the Higgs boson  is emitted
from 
the $Z$ boson is always less than a few per cent at $\sqrt{s}=500$~GeV
 and  can safely be neglected.
In addition, at $\sqrt{s}=500~GeV$, the photon exchange contribution
provides the bulk of the cross section.

The ${\cal O}(\alpha_s)$  inclusive cross section for $e^+e^-\rightarrow t
{\overline t} h$ receives contributions from real gluon emission from
the final quark legs,

\be
e^+ e^-\rightarrow t {\overline t} 
hg\,\,\,\,,
\ee 
 and  also
from virtual gluon contributions to the lowest order process.
 
\begin{figure}[tb]
\centering
\epsfxsize=3.5in
\leavevmode\epsffile{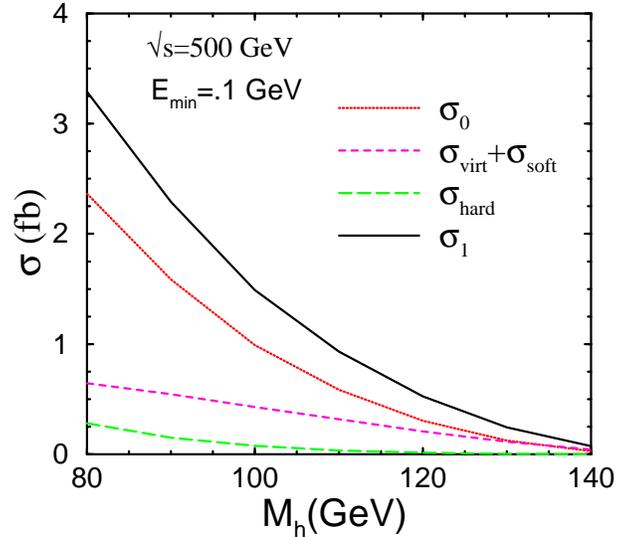}
\caption[]{QCD corrections to $e^+e^-\rightarrow 
t {\overline t}h$ at $\sqrt{s}=500$~GeV.  $\sigma_0$ is
the lowest order cross section and $\sigma_1$ is the
complete ${\cal O}(\alpha_s)$ corrected rate.
We take $M_t=175~GeV$
and $\alpha_s(M_t^2)=.1116.  $} 
\label{sig500fig}
\end{figure}

The real gluon emission is separated into a hard and
a soft contribution by introducing an arbitary cutoff on the gluon
momentum, $E_{min}$.  The infrared divergences in the soft gluon emission
are  then regulated by the introduction of a small gluon mass, $m_g$.
When the one-loop virtual and  the real contributions are combined, the final
result is finite and independent of both $E_{min}$ and $m_g$.  In
Fig.~\ref{sig500fig}, we show the various
contributions to the total cross section.   $\sigma_1$ is the complete
${\cal O}(\alpha_s)$ corrected rate,
\be
\sigma_1=\sigma_0+\sigma_{virt}+\sigma_{hard}+\sigma_{soft}\,\,\,.
\ee
The counterterms are included in $\sigma_{virt}$.  
 The combination $\sigma_{virt}+\sigma_{soft}$ is independent of
the gluon mass, but retains a dependence on $E_{min}$ which is
cancelled by $\sigma_{hard}$.  At $\sqrt{s}=500$~GeV, the corrections
are large and positive, significantly increasing the rate.  The
corrections are smaller at $\sqrt{s}=1$~TeV, with large cancellations
between the hard and the virtual plus soft contributions.
  
\begin{figure}[tb]
\centering
\epsfxsize=3.5in
\leavevmode\epsffile{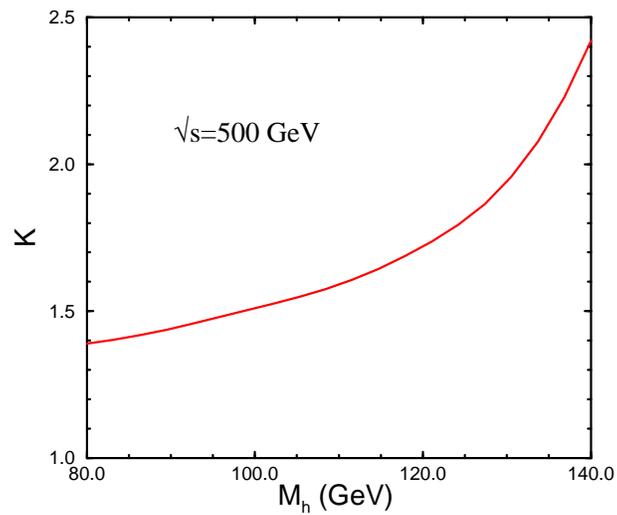}
\caption[]{Ratio of the ${\cal O}(\alpha_s)$ corrected rate to
the lowest order cross section for $e^+e^-\rightarrow t 
{\overline t}h$ at $\sqrt{s}=500$~GeV.
We take $M_t=175~GeV$
and $\alpha_s(M_t^2)=.1116.$}
\label{kfac500fig}
\end{figure}

The size of the QCD corrections can be described by a $K$ factor,

\be
K(\mu)\equiv{\sigma_1\over \sigma_0},
\label{kdef}
\ee

\noindent 
which is shown in Fig.~ \ref{kfac500fig}.
Note that after the cancellation of the ultraviolet
 divergences,
the only $\mu$ dependence is in $\alpha_s(\mu)$.  If $\mu=\sqrt{s}$,
then $K(M_h=100\,\mbox{GeV})$ is reduced to $1.4$ from the value
$K=1.5$ obtained with $\mu=M_t$ for $\sqrt{s}=500~GeV$. 

\section{associated Higgs -top quark
 PRoduction in a supersymmetric model}
\begin{figure}[tb]
\centering
\epsfxsize=3.5in
\leavevmode\epsffile{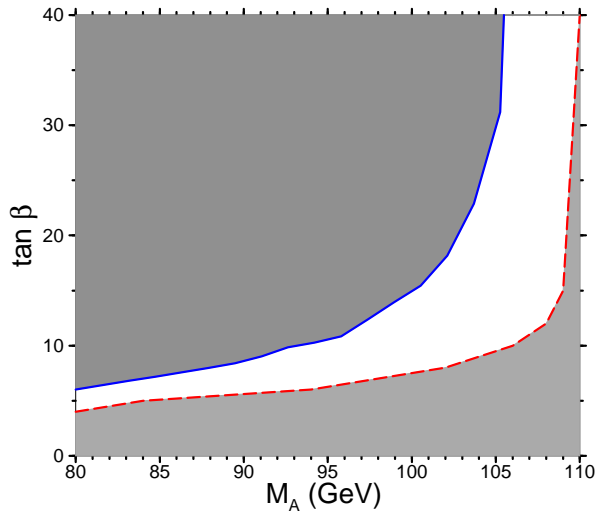}
\caption[]{Regions in the $M_A-\tan\beta$ plane where the
cross section  $e^+e^-\rightarrow 
t {\overline t}h^0$ or
$e^+e^-\rightarrow t {\overline t } H^0$
is larger than $.75~fb$  at $\sqrt{s}=500$~GeV.
 The upper left hand region results from $e^+e^-\rightarrow t {\overline t}
H^0$, while the region at the lower right is the result for
$e^+e^-\rightarrow t {\overline t}h^0$.
All NLO QCD corrections are included.  The squarks are taken to
have a common mass, $M_S=500~GeV$, and
the mixing parameters are set to zero.} 
\label{scanfig}
\end{figure}

 In the minimal supersymmetric model, a top quark pair
can be produced in association with either of the neutral
Higgs bosons, $h_i=h^0,H^0$, or with the pseudoscalar, $A^0$.  The production
of the pseudoscalar is highly suppressed and the
rate for $e^+e^-\rightarrow t {\overline t} A^0$ is less than
$10^{-2}~fb$ at $\sqrt{s}=500~GeV$ for all values of $\tan\beta$
and $M_A$. The rate for either
$e^+e^-\rightarrow t {\overline t}h^0$ or
$e^+e^-\rightarrow t {\overline t}H^0$ is greater than $.75~fb$
throughout most of the $M_A-\tan\beta$ plane and we 
show this in Fig. \ref{scanfig}. 
 We see that this region includes
much of the parameter space. 
The results shown in Fig. 3 are relatively insensitive to changing the 
squark masses
or the mixing parameters of the supersymmetric sector.  

\section{Associated Higgs- bottom quark production in a supersymmetric model}

In the Standard Model,
it will be difficult to extract the bottom quark-Higgs Yukawa coupling from a
measurement of $e^+e^-\rightarrow b {\overline b} h$, since the coupling
itself is tiny and the $Z$ contribution is important, so that there
is a significant dependence on the $ZZh$ coupling. 
In the minimal supersymmetric model,
however,  there are $5$ Higgs bosons,
$\phi=h^0,H^0, A^0, H^\pm$,
so that additional processes not present in the Standard Model may be useful
to pin down the fermion-Higgs boson Yukawa couplings\cite{dr2}.  
In addition,  for
certain values of $\tan\beta$,
the $b{\overline b}\phi$ Yukawa couplings receive
significant enhancements and so the processes  $e^+e^-\rightarrow b {\overline
b}\phi$ may be   larger than in the 
Standard Model.

The physics for $b{\overline b}h^0_i$ production
 is significantly different from that of Higgs production
with a $t {\overline t}$ pair.  In the case of the
$b$ quark,  there is a large
resonant contribution 
 from the process, 
$e^+e^-\rightarrow A^0 h^0_i$,  $A^0\rightarrow
b {\overline b}$.  This enhancement occurs when $M_{h_i}\sim
M_A$ 
and so is relevant for $M_A$ below about $120$~GeV for
$e^+e^-\rightarrow b {\overline b} h_i^0$.

\begin{figure}[t,b]
\centering
\epsfxsize=3.5in
\leavevmode\epsffile{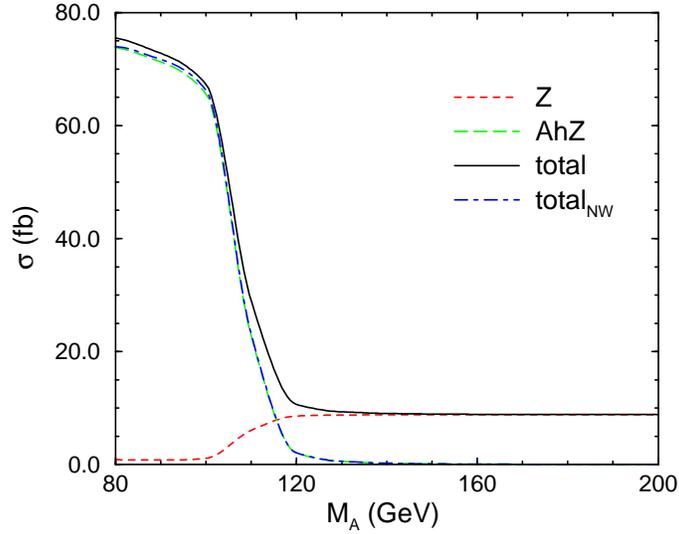}
\caption[]{Contributions to $e^+e^-\rightarrow b {\overline b}h^0$
  at $\sqrt{s}\!=\!500$~GeV for $\tan\beta\!=\!40$.  The
  curve labelled `NW' is the narrow width approximation
to the $A^0$ resonance contribution  
  and includes QCD corrections in the
  resonance region.
The squarks are assumed to have a common mass,
  $M_S\!=\!500$~GeV, and the scalar mixing parameters are set
  to zero.}
\label{fig:bb_lh_tb40}
\end{figure}

Fig. \ref{fig:bb_lh_tb40} shows the different
contributions to the process
 $e^+e^-\rightarrow {b \overline b} h^0$ for 
$\tan\beta=40$ at $\sqrt{s}\!=\!500$~GeV.
The curve labelled ``$total_{NW}$'' is the narrow width approximation
to the  $A^0$ resonance, 
while the curve labelled ``$AhZ$'' is only
 the contribution from the square of the
resonant diagram.  At small $M_A$ ($<120$~GeV), the narrow
width approximation
 is an excellent approximation to the total rate for this
value of $\tan\beta$.  For smaller $\tan\beta$, the narrow width
approximation becomes increasingly inaccurate, since the $Z$ exchange
contribution becomes more and more relevant. At large $M_A$, the rate is
given predominantly by the $Z$ boson exchange contribution and is typically
between $5$ and $10$~fb.

In the narrow width approximation, the QCD corrections to the rate are
trivially included by including the QCD corrections to the
pseudo-scalar width.
 Away from the pseudoscalar resonance,
(large $M_A$), inclusion of the QCD corrections would require a
complete calculation, which we do not include in the present analysis
since the interesting region is near the resonance where the rate is
enhanced.

For heavy Higgs  production, $H^0$,  the narrow width approximation is an
excellent approximation for all values of $\tan\beta$ so the QCD
corrections can be accurately included everywhere.
  For
$\tan\beta<5$, the cross section is larger than $20$~fb even for
$M_A\!\sim\! 200$~GeV.  For $\tan\beta\!>\!5$, the rate is greater than
$20$~fb for $M_A\!>\!110$~GeV, as shown in
Fig. 5.  This process can potentially be used to
probe the couplings of the heavier neutral Higgs boson and to obtain a
precise measurement of the product of the Higgs couplings,
$g_{bbH}g_{ZAH}$.

\begin{figure}[t,b]
\centering
\epsfxsize=3.5in
\leavevmode\epsffile{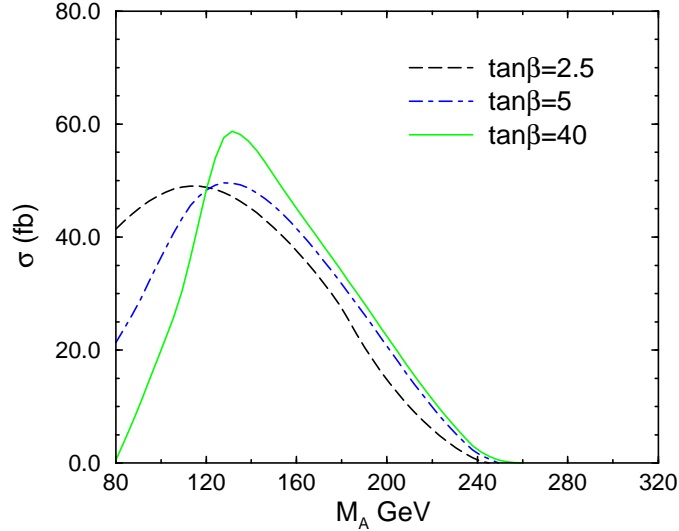}
\caption[]{Cross section for $e^+e^-\rightarrow b {\overline b}H^0$
  at $\sqrt{s}\!=\!500$~GeV including QCD
corrections in the narrow width approximation. 
The squarks are assumed to have a common mass,
  $M_S\!=\!500$~GeV, and the scalar mixing parameters are set
  to zero.}
\label{fddig:bb_lh_tb40}
\end{figure}

\begin{figure}[t,b]
\centering
\epsfxsize=3.5in
\leavevmode\epsffile{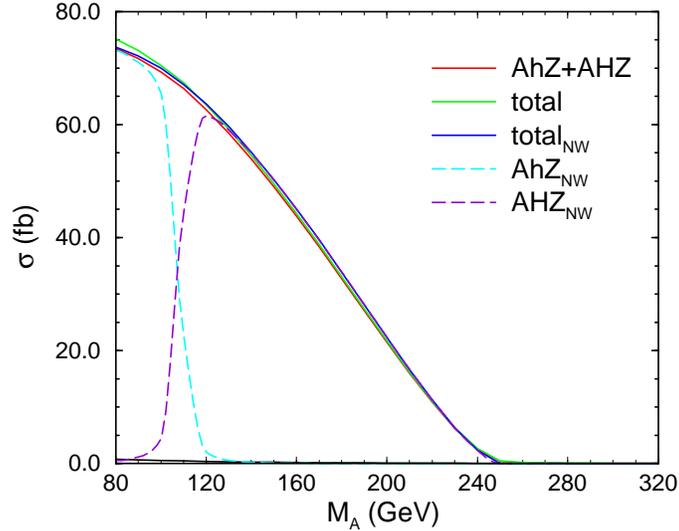}
\caption[]{Total and partial contributions to the
  cross section for $e^+e^-\rightarrow b {\overline b}A^0$ at
  $\sqrt{s}\!=\!500$~GeV at $\tan\beta\!=\!40$, including
  QCD corrections.
The squarks are assumed to have a common mass,
  $M_S\!=\!500$~GeV and the scalar mixing parameters are set
  to zero.}
\label{fig:bba_tb40}
\end{figure}

We can safely work in the narrow width approximation also
for the case of $e^+e^-\rightarrow b {\overline b} A^0$
production.  In fact, in this case the contributions of 
the  $h_i^0$ resonances are
completely dominant and the exact cross section can be
distinguished from the one obtained using the narrow width
approximation only at very high values of $\tan\beta$. The
case $\tan\beta\!=\!40$ is
illustrated in Fig.~
\ref{fig:bba_tb40}. 
Unlike $t{\overline t}A^0$ production, the process
$e^+e^-\rightarrow b {\overline b} A^0$ is not suppressed
relative to $e^+e^-\rightarrow b {\overline b} h_i$
production. For $M_a<200~GeV$, the cross secion is aways greater than $20~fb$.

\section{Conclusion} 
\label{sec:conclusions}

We have computed the ${\cal O}(\alpha_s)$ corrected rate for
$e^+e^-\rightarrow t {\overline t}h_i^0$.  At $\sqrt{s}=500$~GeV, the
corrections are large and positive and this process can
be used to measure the $t {\overline t}h_i^0$ Yukawa couplings,
both in the Standard Model and over much of the parameter space
of the minimal supersymmetric model. 
 In a supersymmetric model, the
rates for $e^+e^-\rightarrow b{\overline b}\phi$ can be
enhanced for large values of $\tan\beta$ and relatively small values
of $M_A$. In such models, the QCD corrections can be accurately
included using the narrow width approximation in the region
where the scalar or pseudoscalar resonance
dominates.   The $b {\overline b} \phi$ 
production processes  will measure a combination of Higgs 
Yukawa couplings.  

\section*{Acknowledgments}
 The work of S.~D. is supported
by the U.S. Department of Energy under contract DE-AC02-76CH00016. The
work of L.~R. is supported by the U.S.  Department of Energy under
contract DE-FG02-95ER40896.

\end{document}